\documentclass[aps,prb,twocolumn,superscriptaddress]{revtex4-2}

\usepackage{graphicx}
\usepackage{multirow}
\usepackage{amsmath,amssymb,amsfonts}
\usepackage{amsthm}
\usepackage{mathrsfs}
\usepackage{appendix}
\usepackage{xcolor}
\usepackage{textcomp}
\usepackage{booktabs}
\usepackage{algorithm}
\usepackage{algorithmicx}
\usepackage{algpseudocode}
\usepackage{listings}
\usepackage{hyperref}
\usepackage{siunitx}
\usepackage[english]{babel}

\begin{document}

\title{Ultrafast dynamics of vibronically dressed core-excitons in graphite: a femtosecond RIXS perspective}

\author{Marco Malvestuto}
\email{marco.malvestuto@elettra.eu}
\affiliation{Elettra Sincrotrone Trieste, SS. 14, Km 163.5, Trieste 34149, Italy}
\affiliation{CNR - Istituto Officina dei Materiali (IOM), SS. 14, Km 163.5, Trieste 34149, Italy}

\author{Beatrice Volpato}
\author{Elena Babici}
\affiliation{Department of Physics, University of Trieste, Via Valerio, Trieste 34140, Italy}

\author{Richa Bhardwaj}
\author{Antonio Caretta}
\author{Simone Laterza}
\author{Fulvio Parmigiani}
\author{Michele Manfredda}
\author{Alberto Simoncig}
\author{Marco Zangrando}
\author{Alexander Demidovich}
\author{Peter Susnjar}
\author{Enrico Massimiliano Allaria}
\author{Alexander Darius Brynes}
\author{David Garzella}
\author{Luca Giannessi}
\author{Primoz Rebernik}
\author{Filippo Sottocorona}
\affiliation{Elettra Sincrotrone Trieste, SS. 14, Km 163.5, Trieste 34149, Italy}

\author{Dino Novko}
\affiliation{Centre for Advanced Laser Techniques, Institute of Physics, Zagreb 10000, Croatia}
\affiliation{Donostia International Physics Center (DIPC), Donostia-San Sebastian 20018, Spain}

\begin{abstract}
This study demonstrates one of the first implementations of time-resolved resonant inelastic X-ray scattering (tr-RIXS), marking a seminal extension of RIXS spectroscopy into the ultrafast time domain. By investigating the ultrafast dynamics of vibronically dressed core excitons in graphite using femtosecond X-ray pulses from a Free Electron Laser, we reveal previously inaccessible insights into the transient coupling between core excitons and specific optical phonon modes. Our approach establishes tr-RIXS as a powerful, transformative tool capable of elucidating the intricate interplay between electronic and lattice dynamics, opening new avenues in ultrafast materials research.
\end{abstract}

\maketitle

\section{Introduction}
Excitons, defined as Coulomb-bound electron-hole pairs, play a fundamental role in understanding various phenomena in condensed matter physics. These states emerge across a wide range of materials, including semiconductors, molecules, semimetals, low-dimensional systems, and metals with weak electronic screening. \cite{YuCardona.2005}. Among excitonic subclasses, core excitons (CE) remain relatively unexplored. These excitons form when a photon excites an electron from a core atomic level into the conduction band, resulting in a strongly localized excitonic state. Despite their typically short lifetimes ($\leq10$ fs, depending on the core level and atomic species), CEs significantly impact the scattering processes in materials where exciton screening is ineffective.

A key manifestation of CE effects is the renormalization of X-ray absorption spectra (XAS). At the carbon K-edge, XAS provides critical insights into the electronic properties of molecules and graphitic systems\cite{Zhang.2012alc,Wessely.2005,Veenendaal.1997,Ahuja.1996,Bruhwiler.1995,Skytt.1994,Ma.19938f,Sette.1990,Weng.1989,Mele.1979}. However, the spectra are shaped by intermediate excitonic states, necessitating advanced theoretical modeling for accurate interpretation. Furthermore, CEs significantly influence the structure of resonant inelastic X-ray scattering (RIXS) spectra. Recent research demonstrates that CEs can activate optical phonons \cite{Feng.2020,Gilmore.2022,Dashwood.2021,Geondzhian.2020,Harada.20041d,Ma.19938f}, emphasizing the dynamic coupling between excitons and phonons. This coupling is particularly critical in time-resolved RIXS, where photoinduced screening of CEs and the excitation of strongly coupled hot optical phonons alter the spectroscopic fingerprints of exciton-phonon interactions.

The electronic screening of core holes, a complex and pivotal aspect of radiation-matter interactions, determines the spectral features in both XAS and RIXS. Although metals efficiently screen core holes because of their high density of conduction electrons, leading to negligible excitonic effects, materials with reduced screening capabilities can sustain bound excitonic states. CEs may distort the local crystal lattice in such systems, resulting in vibronically dressed excitons that enhance phonon activation.

Graphite, an archetypal two-dimensional material with pronounced electronic anisotropy, provides an exemplary case for studying CE-phonon interactions \cite{Ma.19938f,Attekum.1979,Ahuja.1996,Veenendaal.1997}. Experimental and theoretical investigations have shown that carbon 1s core holes in graphite, which are inadequately screened in this van der Waals system, trigger multiple scatterings of CEs with optical phonons. These phonons are activated by local Jahn-Teller (JT) distortions caused by the unscreened core hole.

In this study, our objective was to modulate the screening efficiency of core holes in graphite, induced by femtosecond extreme ultraviolet (XUV) pulses, and to monitor the temporal evolution of the vibronic dressing of CEs. By optically exciting the system to generate an electron-hole plasma in the conduction and valence bands, we increased the available charge density in this semimetal, enhancing the screening capability. The extent of screening was indirectly probed through the dynamics of the phonon sidebands in the RIXS loss spectrum, serving as a quantitative indicator of exciton-phonon coupling strength.

\begin{figure*}
    \centering
    \includegraphics[width=0.9\linewidth, scale=0.9]{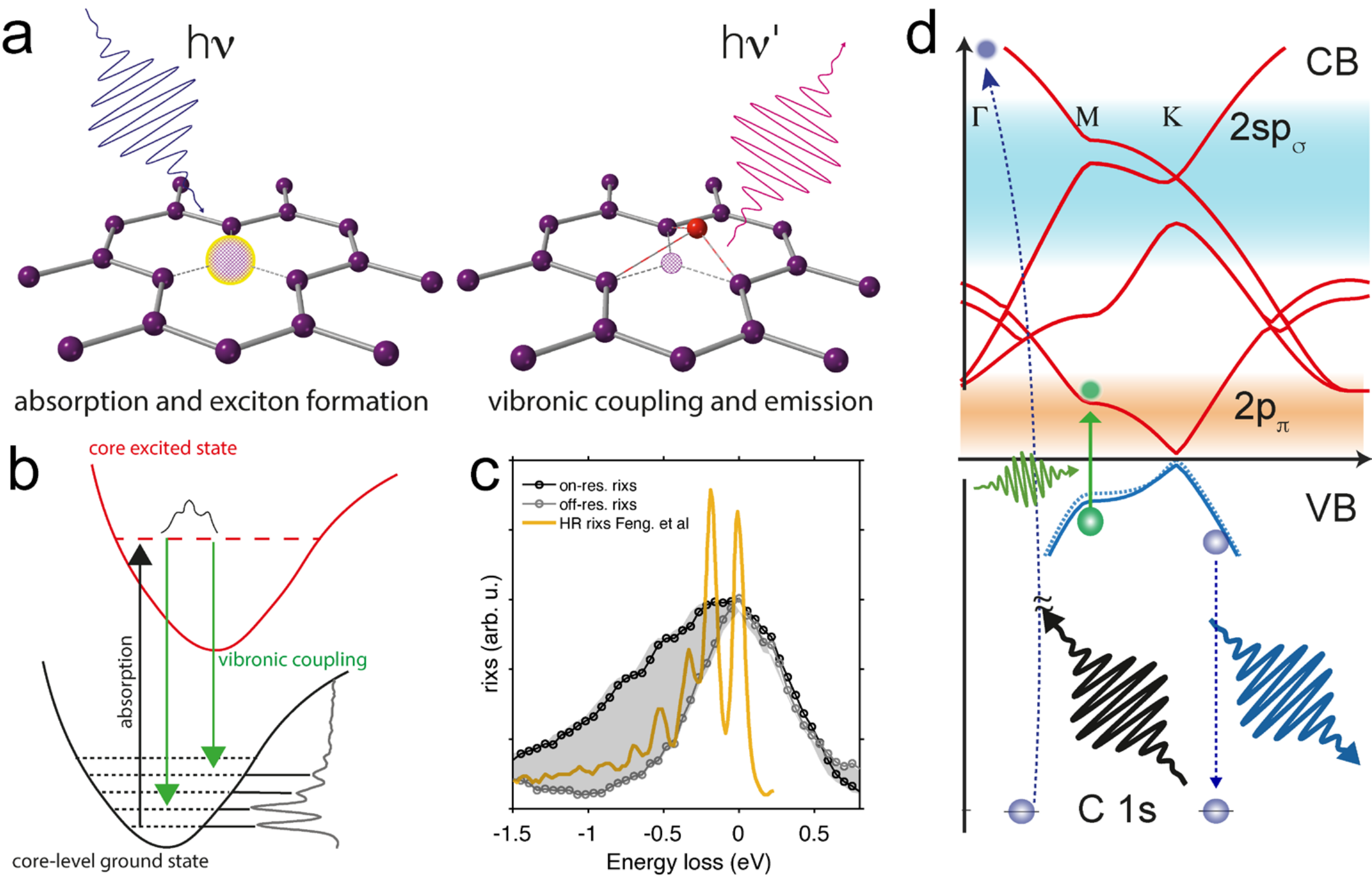}
    \caption{panel \textbf{a}: an FEL pulse, tuned to the carbon absorption edge, excites the system by creating a core-hole exciton which is strongly coupled to mode-selective phonons. After a characteristic scattering time duration, the system emits a photon with altered energy. panel \textbf{b}: a schematic illustration of the RIXS scattering process leading to phonon excitation. The vibrational wavefunction in the ground state projects onto multiple vibrational wavefunctions in the excited state, resulting in the formation of a core-hole exciton. Upon decay, the excited-state vibrational wavepacket projects back onto multiple ground-state vibrational levels, leading to a RIXS loss spectrum that features a harmonic progression of phonons. Panel \textbf{c}: example of high-resolution RIXS data (yellow curve)\cite{Feng.2020} alongside data from the present experiment measured on resonance (black open dots) and off-resonance (gray open dots), highlighting resonant overtones threshold as a function of energy loss. The shaded area is the difference between on- and -off resonance spectra and represents the loss tail. Panel \textbf{d}: the optical pump pulse, tuned to the resonant energy of the \(\pi-\pi^*\) transition at the \(M\) saddle point of the Brillouin zone (BZ), generates an electron-hole plasma confined between the graphene layers. Due to the BZ topology, a bottleneck effect extends the average lifetime of valence band quasiparticles \cite{Ishioka.2008, Spataru.2001, Pagliara.2011}. Subsequently, an X-ray probe pulse induces the formation of core excitons through the transition of an electron from the core shell to the conduction band $2sp^2$ orbitals with \(\sigma\) symmetry. The excited state then decays back to the ground state within a defined effective scattering time \(\tau_s\) \cite{Williams.1974,Gelmukhanov.1994}.
    }
    \label{figure1}
\end{figure*}
\section{Experimental}

A conceptual scheme of the RIXS process in graphite is depicted in Figure \ref{figure1}a-b.
An incident X-ray photon promotes an electron from the 1s core orbital in the ground state to the \(2sp^2_{\sigma}\) bands, where it experiences a partially screened potential caused by the core hole. 
The presence of the core hole in the intermediate state induces a substantial modification in the local atomic configuration around the interaction site \cite{Ament.2011, Carlisle.1994, Gelmukhanov.1999, Feng.2020, Yavas.2010, Veenendaal.97, Harada.2004ab}. This modification arises from the ionic coupling between the core exciton and lattice phonon systems, leading to dynamical symmetry breaking within the core-excited states. As a result, a significant JT distortion ($\sim$ 0.20 \(\si{\angstrom}\)) occurs, displacing the core-excited atom from its equilibrium position in the ground state. 

The excited electron subsequently relaxes in a few femtoseconds (the inverse of the natural linewidth $\Gamma$ associated with the core-hole state) to fill the core hole, emitting a photon in the process. 
The energy difference between the incident and scattered photons manifests as overtones corresponding to the energy of the phonons \(\hbar\omega_{\text{p}}\) remaining in the final state as depicted in Figure \ref{figure1}c.

Representative static on resonance (black curve) and off-resonance (gray curve) RIXS loss spectra obtained in this experiment, with an energy resolution of approximately 350 meV, are presented in Figure \ref{figure1}c. For comparison, a high-resolution RIXS loss spectrum, reported by Feng et al. \cite{Feng.2020}, with an energy resolution of approximately 50 meV and measured at a similar incident photon energy is also shown.
The high-resolution spectrum features a prominent elastic peak at 0 eV energy loss, accompanied by a series of vibrational replicas (overtones), equally separated in energy corresponding to the \( E_{2g} \) phonon energy ($\sim$196 meV) \cite{Feng.2020, Gelmukhanov.1994}. Thus, each replica represents multiple energy losses with an increasing number of phonons\cite{Dashwood.2021,Geondzhian.2018}. 
In the present experiment, individual vibronic overtones are not resolved because of the limited energy resolution. However, a broad phonon sideband, reminiscent of those observed in previous studies by Ma et al. \cite{Ma.19938f} and Harada et al. \cite{Harada.20041d}, is evident in the resonance spectrum. In Feng et al., the much less intense overtones from the $A_{1g}$ mode are also resolved but are neglected in the present experiment.

The experiment was carried out at the time-resolved RIXS endstation of the MagneDyn beamline \cite{Malvestuto.2022} at the FERMI Free Electron Laser, Elettra Sincrotrone Trieste, Italy. 
A standard pump-probe experimental setup was used, as described in reference \cite{Malvestuto.2022}.
The highly oriented pyrolytic graphite (HOPG) sample was cleaved in air prior to its placement in the vacuum chamber.

An FEL pulse, with the photon energy tuned across the carbon K edge (\(1s \rightarrow 2sp^2_\sigma\)), was used to drive the inelastic scattering process. The energy resolution of the source corresponds to the bandwidth of the pulse, approximately 60 meV. The polarization of the FEL pulses was maintained in the horizontal plane of scattering, and the beam incidence angle was quasi-collinear to the pump beam, i.e. 30$^\circ$ relative to the normal incidence direction. These conditions were chosen to maximize the cross-section at the K\(_\sigma\) threshold (\(1s \rightarrow 2sp^2_\sigma\)), ensuring a significant component of the polarization in the \(x-y\) plane of the sample.
The resonantly emitted pulses were captured by the spectrometer on a stroboscopic, pulse-to-pulse basis.
Inelastic spectra were acquired using a Scienta XES355 soft X-ray spectrometer \cite{Nordgren.2000erp}. 

The third harmonic (263 nm, 4.71 eV, p-polarization, 50 Hz repetition rate) of the fundamental 795 nm FERMI User laser \cite{Danailov.2014, Sigalotti.2013,Malvestuto.2022} was used to resonantly excite the system at the \(\pi - \pi^{*}\) high-symmetry M point in the Brillouin zone (BZ) of graphite (see Figure \ref{figure1}d). Each optical pump pulse had an intensity of 10 microjoules. Considering the pump spot size and an incidence angle of 32.5 degrees with respect to the normal to the sample, the resulting optical fluence on the sample was about 32 mJ/cm\(^2\), corresponding to approximately 14\% of excited atoms (see SI). 

The selection of an optical pump resonating at the saddle point \(M\) of the Brillouin zone is based on two considerations. The first consideration is that this configuration ensures that the optically excited orbitals (2p\(_z\)) remain decoupled from those directly probed by the core transition \(1s-2sp^2_\sigma\), thus avoiding saturation effects. This decoupling mitigates spurious influences on the core absorption process, preventing phenomena like XAS transparency due to saturation of the absorption channel or renormalization of the XAS lineshape.

Secondly, when valence quasiparticles are resonantly excited at the M saddle point, the resulting hot carriers (\(\pi - \pi^*\)) exhibit a lifetime that exceeds the temporal resolution of the present experiment ($\sim$50 fs) and surpasses the typical lifetime of the carbon core exciton (\(\sim\)10 fs). The anomalous behaviour of valence quasiparticle lifetimes in graphite has been extensively documented in the literature\cite{Xu.1996, Breusing.2011, Spataru.2001, Piscanec.2004, Park.2009, Park.2007}. This behaviour is attributed to the phonon bottleneck effect\cite{Johannsen.2013, Caruso.2020, Novko.2019}, which arises from the distinctive band structure of graphite. Specifically, this effect leads to significant variations in quasiparticle lifetimes depending on the wave vector direction within the Brillouin zone (BZ).

In contrast, when hot carriers are generated by optical pulses with energies near the K point in graphite, their lifetimes are typically much shorter, on the order of tens of femtoseconds. This rapid decay is due to efficient carrier-carrier and carrier-phonon scattering processes\cite{Duvel.2022, Girotto.2023, Sidiropoulos.2021, Johannsen.2013, Breusing.2008, Breusing.2011, Kampfrath.2005}.

\section{Results and discussion}

\begin{figure*}
    \centering
    \includegraphics[width=0.9\linewidth, scale=0.9]{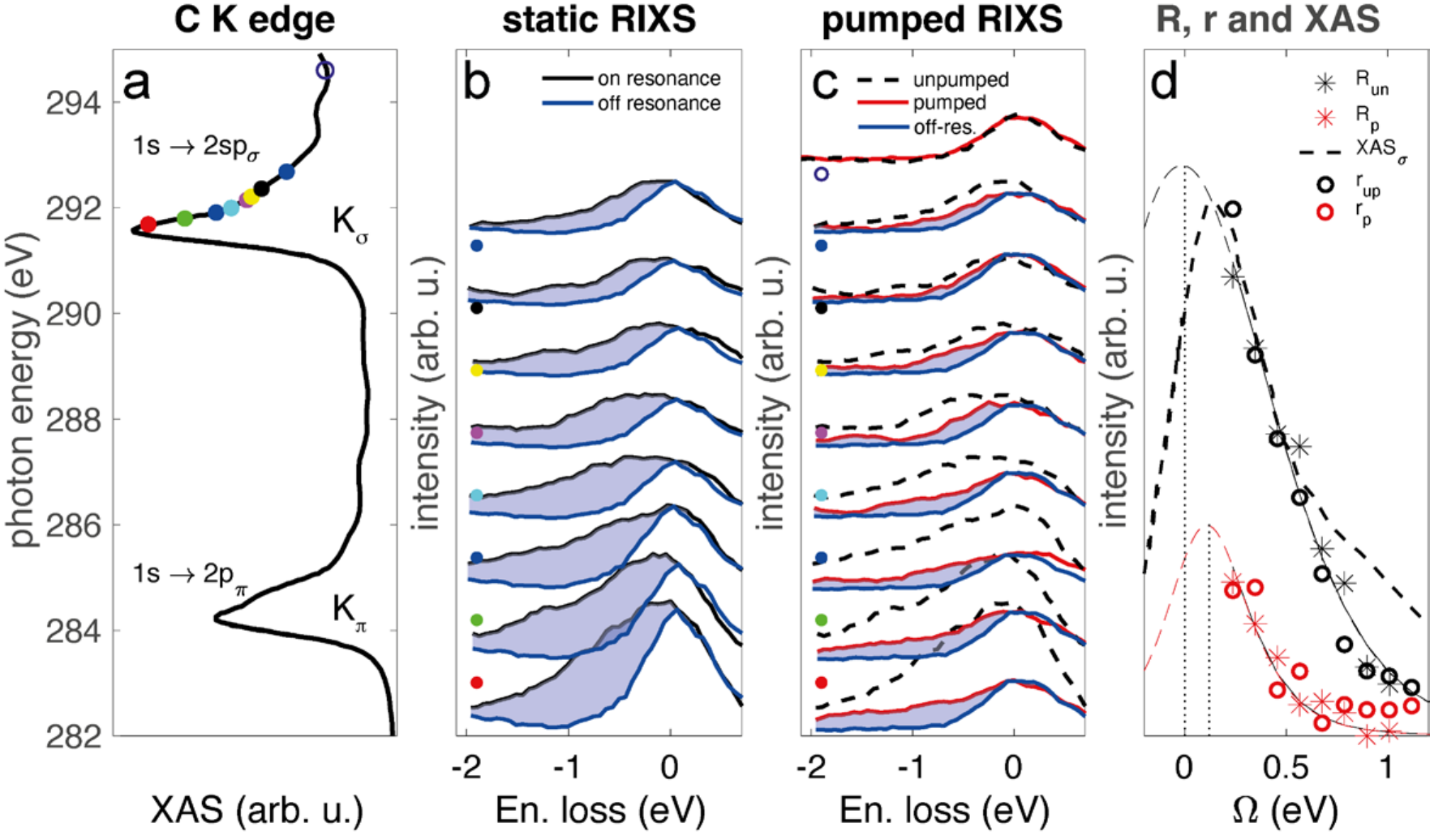}
    \caption{Panel \textbf{a}: C K-edge XAS of HOPG. The energy of the FEL photon pulse is varied across the \(1s-2sp^2_{\sigma}\) channel. Panel \textbf{b}: unpumped RIXS loss spectra (black curves) recorded at different on-resonance photon energies (indicated by coloured dots in panel a) across the carbon K-edge, plotted as a function of energy loss. These on-resonance spectra are compared with an off-resonance unpumped spectrum (blue curves). The maximum intensity of the off-resonance spectrum is rescaled by a factor \textbf{r} to match the intensities of the resonant spectra at the same energy loss. The blue-shaded area between the on- and off-resonant spectra represents the intensity distribution of overtones (\textbf{R} in the text and panel \textbf{d}), with the intensity of this area serving as a measure of exciton-phonon coupling. Panel \textbf{c}: similarly to panel b, the resonant pumped (red curves) inelastic spectra, recorded at both resonant and off-resonant photon energies, are displayed. For comparison, the unpumped spectra (dashed black curves) are also displayed. At off-resonant photon energies, the pump effect is absent, and the phonon tail vanishes from the lineshape, yielding a symmetric profile (top pair of curves). Panel \textbf{d}: the integrated inelastic spectral weights \textbf{R}  for the pumped (red star dots) and unpumped (black star dots) spectra are plotted as a function of the detuning energy and compared with the carbon K\(_{\sigma}\) XAS spectrum (black dashed curve) and the coherent spectral weight factor \textbf{r}. The theoretical intensity distribution of the overtones is also included for comparison (red and gray dashed curves, see text and equation \ref{eq:intensity}).}
    \label{figure2}
\end{figure*}
The absorption spectrum of graphite is shown in Figure \ref{figure2}a at both the \(\pi\) and \(\sigma\) thresholds.  The peak at the lower photon energy ($\sim284$ eV) corresponds to the transition of the 1s photoelectron into the unoccupied out-of-plane 2p\(_{\pi^*}\) orbitals (1s \(\rightarrow \pi^*\)). Conversely, the peak at a higher energy ($\sim291.8$ eV) is assigned to contributions from the hybridized 2sp\(^2_\sigma\) in-plane \(\sigma^*(\text{C–C})\) orbitals.
This higher energy edge does not directly reflect the electronic dynamics induced by the photoexcitation of the \(\pi \rightarrow \pi^*\) transition. Instead, it primarily responds to structural changes within the material's configuration\cite{Sidiropoulos.2021}.
The graphite XAS is predominantly influenced by Frenkel-type excitonic effects at resonance energies \cite{Mele.1979, Ritsko.1982, Ahuja.1996, Wessely.2005}. 
The one-electron density of states alone does not explain the absorption spectra because the excited electron occupies a highly localized state pulled off the bottom of the band gap. This mechanism leads to the appearance of white lines and a significant modification of the cross-section \cite{Carlisle.2000}. Similar effects are evident in the resonant emission spectra of graphite at the carbon \(K\) edge \cite{Skytt.1994, Skytt.1995, Skytt.1996, Ma.1992, Ma.19938f, Bruhwiler.1995}.
 
In Figures \ref{figure2}\textbf{b-c}, the experimental data illustrate the impact of core-exciton vibronic interactions on the loss spectra in both the ground and the optically excited states. 
Figure \ref{figure2}\textbf{b} shows the unpumped loss spectra of graphite (black curves), recorded at FEL photon energies (indicated by coloured markers in panel \textbf{a}) tuned across the \(K_{\sigma}\) resonance. These spectra are plotted as a function of energy loss and displayed in order of increasing photon energies. The on-resonance spectra are compared with an off-resonance unpumped spectrum (blue curve), recorded at a photon energy beyond the absorption edge (blue open circle in panel a).
As aforementioned, when the photon energy $h \nu_{\text{in}}$ is tuned close to the absorption resonance ($h\nu_{\text{res}}$\(\sim 291.8\) eV), thereby reducing the energy detuning value \(\Omega = h(\nu_{\text{in}} - \nu_{\text{res}})\), an asymmetric incoherent loss shoulder emerges, as indicated by the colored areas between the blue and black curves in figure \ref{figure2}b.
The lineshape of the off-resonance spectrum lacks the asymmetric loss tail and exhibits a symmetric profile. This symmetry arises because, in the emission process, the final state of the RIXS process mirrors the initial state and does not involve exciton-phonon coupling scattering. This portion of the spectrum is commonly referred to as the coherent component.
The intensity of the off-resonance spectrum has been rescaled so that its maximum at zero energy loss matches the intensity of the corresponding on-resonance spectrum. This rescaling enables a distinction in the on-resonance spectra between the coherent scattering contribution and the inelastic component originating from the overtone distribution. The pumped and unpumped rescaling factors \textbf{r} are shown in panel d of Figure \ref{figure2} (black and red open circles) as a function of \(\Omega\). 

The decrease in intensities of the loss part of the spectra with \(\Omega \) can be attributed to the shorter effective scattering time \(\tau_s=\frac{1}{\sqrt{\Gamma^2 + \Omega^2}}\), where \(\Gamma \) is the excitation lifetime (see for example \cite{Gelmukhanov.1997, Ament.2011,Geondzhian.2020}), which suppresses the RIXS channel that involves interactions with slower degrees of freedom, such as lattice vibrations. 
The concept of effective scattering time is introduced and discussed by Gel'mukanov\cite{Gelmukhanov.1994}, for which a summary is given in SI.

Panel c of Figure \ref{figure2} presents pairs of pumped (red curve) and unpumped (black curve) RIXS loss spectra as a function of energy loss, measured at the same photon energies as those in panel b. The delay between the pump and probe pulses for the excited spectra was set at 150 fs. The unpumped spectra (dashed black curves) are also plotted for comparison. For spectra measured with small detuning energies from the resonance, the intensity of the loss tail is noticeably reduced. The loss spectral weight of the spectrum is calculated as 
\[
\textbf{R}_{p/up} = \textbf{I}_{p/up}^{\text{on}} - \textbf{I}_{p/up}^{\text{off}}
\]
where \textbf{I} denotes the intensities of the pumped (p)/unpumped (up) on- and off-resonance spectra integrated in the energy-loss range [-2:0] eV. 

Particular attention should be given to the pair of pumped and unpumped spectra measured off-resonance (top spectra in panel c of Figure \ref{figure2}, blue open dot), specifically at a photon energy of 294.5 eV with a detuning of approximately \(\Omega = 2.7\) eV. In this case, the overtone tails are absent due to the large detuning value and the spectra are symmetric. Furthermore, the effect of the pump on the excited spectrum is negligible, as indicated by the complete overlap of the pumped and unpumped spectra.

This lack of pump-induced changes in the off-resonance spectra is a key observation, highlighting that the influence of the photoinduced plasma predominantly impacts the excitonic state and its dynamics. Properties such as exciton energy, screening, and vibronic coupling are affected, while the density of states remains largely unchanged in the out-of-equilibrium state.

The integrated intensities \textbf{R} of the pumped and unpumped spectral losses (red and black star markers), are plotted as a function of the photon energy detuning $\Omega$ in Figure \ref{figure2}d. 
A significant decrease in intensity is observed for the excited system compared to that of the unexcited system. Additionally, the difference between the two intensities diminishes and eventually disappears as the detuning of the photon energy relative to the threshold increases. This trend confirms a diminishing impact of the photoexcitation on the system as the photon energy deviates from the resonance energy.

We used the simplified single-site framework developed by Ament et al. \cite{Ament.2011} to model the intensities \(\textbf{R}\) of pumped and unpumped phonon sidebands as a function of energy detuning. This model approximates the system by considering a single isolated site where the local electron density in the valence orbital couples with lattice displacements. Although simplified, this approach effectively predicts the series of low-energy harmonic excitations in the energy-loss spectra and correlates their relative intensities with the strength of the electron-phonon interaction. To extract the electron-phonon coupling (EPC) strength from the experimental RIXS data, a fitting procedure was conducted making use of the phonon intensity formula as a function of the detuning parameter and following the approach already used in previous works \cite{Rossi.2019, Feng.2020, Ament.2011, Gelmukhanov.2021}. The relevant equation, which underpins the fitting process, is as follows:

\begin{equation}
\textbf{R}= \frac{e^{-2g}}{g} \left| \sum_{n=0}^{n_{\text{max}}}\frac{g^n (n - g)}{n!(\Omega + i\Gamma + (g - n)\omega_0)} \right|^2 \,
\label{eq:intensity}
\end{equation}

In the expression, the summation accounts for the series of phonon excitations, \( g \) represents the dimensionless EPC parameter, \(\Omega\) the detuning energy, \(\Gamma\) the natural energy broadening of the core hole, and \(\omega_0\) is the frequency of the scattering phonon. 
In this study, a single optical phonon mode (\(E_{2g}\) G mode at the \(\Gamma\) point) coupled to the exciton was assumed, with the total loss intensity \(\textbf{R}\) used as a measure of the EPC. Additionally, the core lifetime \(\Gamma\) was assumed to remain unchanged in the out-of-equilibrium state.

For each experimental data set, the fitting procedure (see SI) involved optimizing the EPC strength \(g\) to achieve the best agreement between the experimental results and the theoretical model. The unpumped data were used as a reference baseline, providing the intrinsic EPC strength characteristic of the system in equilibrium. In contrast, the pumped data, recorded following photoexcitation, exhibited a reduced EPC strength, reflecting the dynamic modifications induced by the pump pulse.

The fitting functions $I_p$ and $I_{up}$ (red and black dashed curves) are compared with the experimental data in panel d. The reference value for the static state, \(g_{up} = 4.55\pm1.1\) and \(M_{up} = \sqrt{g}\omega_0 = 0.42~eV\), aligns well with the values reported in the literature for graphite\cite{Feng.2020}. In contrast, the values obtained for the excited state are \(g_{p} = 0.32\pm0.6\) and \(M_p = 0.12~eV\).
The interpretation of this reduction in the electron-phonon coupling parameters is discussed in the following section. 

The effects of the optical pump on the parameters \(\textbf{R}\) (representing the inelastic component of the spectra) and \(\textbf{r}\) (the rescaling factor representing the coherent component, see caption of Figure \ref{figure2}), along with the fitting results, suggest the following scenario. At a fixed delay of 150 fs between the pump and probe pulses, the coherent component \(\textbf{r}\) decreases compared to the unpumped case. Since \(\textbf{r}\) reflects changes in the absorption cross-section, its reduction indicates a decrease in the white line intensity characteristic of the resonant K-edge XAS spectrum. As the white line directly results from Frenkel core-exciton formation, this reduction provides initial evidence of core-exciton screening.

The fitting results (Figure \ref{figure2}d) further reveal a positive energy shift of approximately 120 meV in the maximum of the fitted function for the pumped spectrum (indicated by vertical lines in panel d). This energy shift can be interpreted as a change in the absorption resonance, offering additional evidence of exciton screening. The excess charge generated by the optical pump effectively screens the core hole compared to the unpumped case, reducing the exciton binding energy. Consequently, the core-bound electron occupies less localised states, which shift toward the lower-energy region of the band gap in the one-electron density of states.

For the parameter \(\textbf{R}\), the fitting results indicate that the EPC strength decreases in the non-equilibrium state.  
Interestingly, the similar detuning dependence of both \(\textbf{R}\) and \(\textbf{r}\) suggests that core-hole screening by the hot electron-hole plasma and energy transfer from the plasma to the phononic system occur simultaneously. Johannsen et al. \cite{Johannsen.2013} have shown that ultrafast electron dynamics in graphene, and similarly in graphite, involves rapid electron-electron scattering and thermalization of the electronic subsystem. This aligns with studies reporting sub-picosecond energy transfer to strongly coupled optical phonons, followed by energy dissipation through optical phonon emission \cite{Caruso.2020, Yang.2017uac}. These timescales agree well with the evolution of the phonon sideband spectral weight observed in the present study.

\begin{figure*}
    \centering
    \includegraphics[width=0.95\linewidth, scale=0.9]{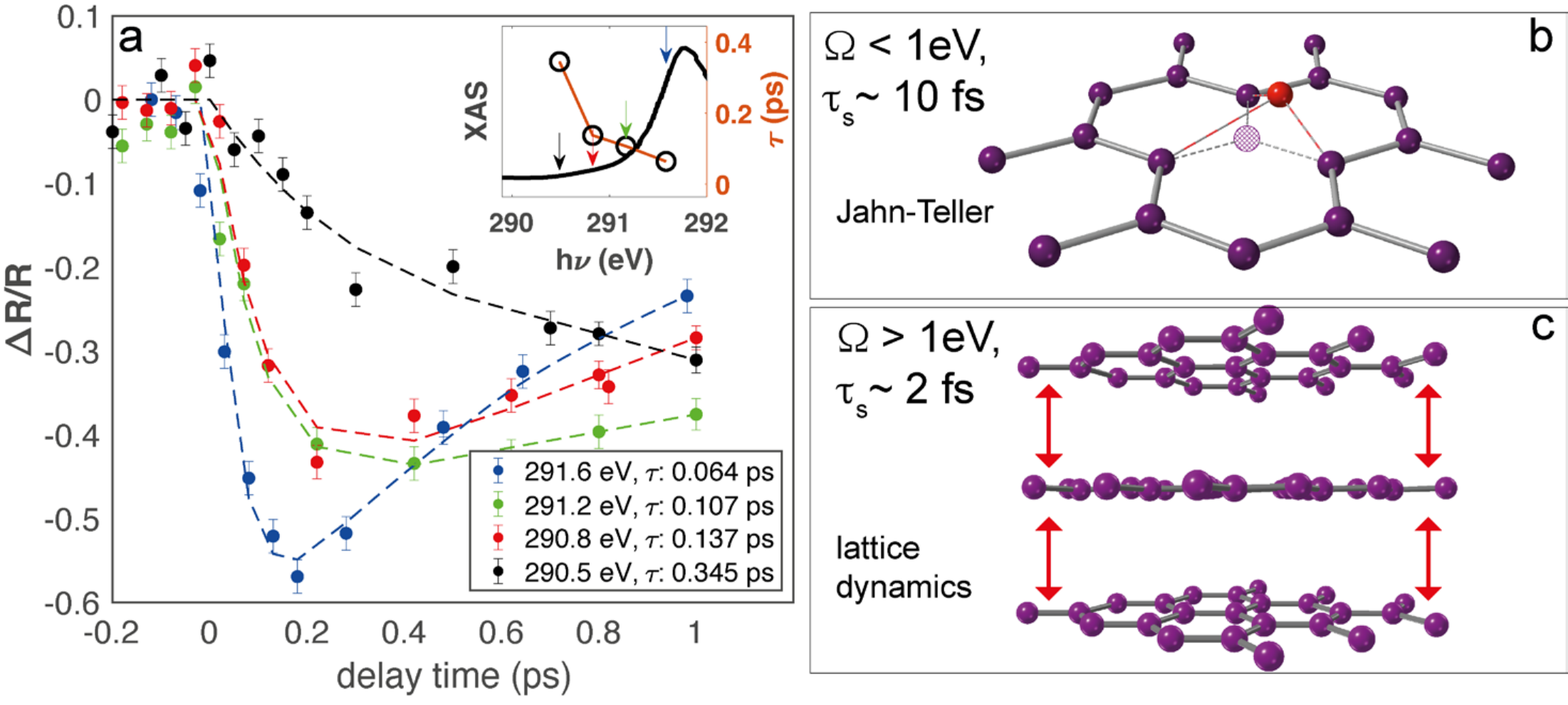}
    \caption{Panel \textbf{a}: dynamics of the relative change  in the integrated intensity of overtones \(\mathbf{R}\) as a function of the time delay. The data were measured at four different incident photon energies varying detuning relative to the absorption threshold. The inset shows the carbon $K_{\sigma}$ threshold, with coloured arrows indicating the photon energies at which the dynamics displayed in the main figure were measured. The red line and markers represent the fitted decay times. Panels \textbf{b-c}: the two dynamic regimes observed in the temporal traces correspond to distinct physical mechanisms. For small detuning energies with an effective scattering time \(\sim\) 10 fs, the rapid decrease in inelastic spectral weight reflects reduced exciton-optical phonon scattering, attributed to the screening effect of the core-exciton by the photoinduced electron-hole plasma. In this regime, the core hole is more effectively screened by the interlayer electron-hole plasma, leading to a reduction in both the local JT distortion and the core exciton-optical phonon coupling. In contrast, when the detuning \(\Omega\) exceeds 1 eV (effective scattering time \(\leq\) 2 fs), the observed slower dynamics are attributed to thermal modifications of the crystal lattice.}
    \label{figure3}
\end{figure*}
The temporal evolutions of the relative change of \textbf{\(R\)} (\textbf{\(\Delta R/R\)}) are plotted in the main panel of Figure \ref{figure3}. These dynamics were recorded at four different FEL photon energies across the resonance threshold, as indicated by the arrows in the inset of Figure \ref{figure3}a. The analysis of the dynamic traces was carried out using a model double exponential decay-recovery function, as outlined in the SI section, with a rapid decay time followed by a slower recovery process. The values corresponding to these ultrafast dynamic processes are plotted (red open circles) in the inset of Figure \ref{figure3}a.

The time traces reveal two distinct regimes as a function of the energy detuning \(\Omega\) governed by different physical mechanisms. However, the transition between these regimes is not defined by a precise value of \(\Omega\) but rather by its extremes, when \(\Omega\) approaches the resonance threshold or deviates significantly from it. As a result, the two regimes coexist, with their relative contributions to the \(\Delta R/R\)  dynamics varying continuously with \(\Omega\).

For small \(\Omega\), the effective scattering time \(\tau_s\) is approximately 10 fs. In this regime, a rapid decrease in the inelastic sideband spectral weight with a time constant of approximately 65 fs is observed, approaching the experimental temporal resolution. This behaviour is driven by an efficient screening of the core exciton by the photoinduced electron-hole plasma, which reduces the exciton-optical phonon interactions and suppresses the local JT distortion (see Figure \ref{figure3}b). The resulting dynamics reflects a weakening of the overall exciton-phonon coupling, consistent with findings from transient absorption experiments (Sidiropoulos et al.\cite{Sidiropoulos.2021}) highlighting the role of excited carriers, particularly in the energy region dominated by \(\pi\) states.

In contrast, when the energy detuning \(\Omega\) exceeds 1.0 eV the effective scattering time decreases to below 2 fs (as illustrated in Figure 2S and discussed in SI A.2) and the fast local motion becomes insignificant in the scattering processes. This occurs because the onset times necessary for the local atomic structural motion exceed the effective scattering time (see Figure \ref{figure3}\textbf{c}). 
Within this regime, the dynamics exhibits a characteristic time \(\tau\) of $\sim$~330 fs, which is attributed to thermal modifications in the interplanar distances, specifically involving the adiabatic contraction and expansion of the crystalline structure. The observed dynamics aligns well with previous studies using time-resolved structural probes, which have reported similar adiabatic responses in the crystalline structure of graphite under picosecond optical pulse excitations \cite{Harb.2011, Carbone.2008, Carbone.2010}. 

The roles of exciton screening and phonon coupling provide crucial insight into the observed dynamics. When the detuning is small, the coupling between excitons and mode-selected optical phonons dominates the transient response. The enhanced coupling between the phonon mode and the \(\sigma\)-resonance exciton is not merely due to stronger local perturbation by the exciton compared to delocalised particle-hole pairs probed in other experimental techniques. Instead, this difference arises from the distinct electronic states accessed by RIXS. When resonant at the $K_{\sigma}$ edge, RIXS probes the \(\sigma\)-resonance states, which exhibit unique symmetry properties, matrix elements, and phase space configurations. These states fundamentally differ from the low-energy \(\pi\) and \(\pi^*\) states around the Fermi level, primarily studied in techniques like trARPES, transport measurements, and IR optical conductivity. In these methods, which are sensitive to Fermi surface dynamics, coupling predominantly involves the \(\pi\) and \(\pi^*\) states and favours the zone-edge A\(_1'\) phonon mode at the K point. In contrast, RIXS involves an intermediate core-excited state at the \(\sigma\)-resonance, which couples more strongly to the \(E_{2g}\) phonon at the \(\Gamma\) point. The orbital-resolved nature of this coupling is well supported by theoretical calculations, as described in the SI \ref{supp-method:calc}.

At larger detuning, the observed dynamics suggest the involvement of out-of-plane \(\pi^*\)-state interactions, as phonons less coupled to the core-exciton come into play. The interplay between these modes and the crystalline lattice leads to thermally driven structural changes, including adiabatic interplanar contractions and expansions, consistent with previous reports of lattice responses under similar excitation conditions \cite{Novko.2019,Carbone.2010}.

\section{Conclusions}

This study sheds light on the intricate mechanisms by which photoinjected charges in the valence and conduction bands influence the screening dynamics of core excitons in graphite, particularly their coupling with optical phonons. By employing time-resolved resonant inelastic X-ray scattering (RIXS) combined with photon energy detuning, we accessed physical processes occurring on femtosecond timescales, exceeding the temporal resolution of the experimental setup. Our findings reveal that both the photoinduced screening of core excitons by an electron-hole plasma and the initiation of Jahn-Teller distortions occur with ultrafast dynamics, leaving clear spectroscopic signatures in this prototypical van der Waals material.

The chemical and orbital selectivity of RIXS underscores the unique role of \(\sigma^*\)-state interactions under resonant conditions and their robust coupling to the \(E_{2g}\) optical phonon mode. This strong coupling governs not only the timescales of electron-phonon scattering but also the modulation of inelastic sideband intensities. These observations highlight the critical role of mode-selective phonon dynamics in shaping the RIXS spectral features of graphite.

A pivotal outcome of this study is the identification of two distinct dynamic regimes, each governed by separate physical mechanisms, depending on the photon energy detuning \(\Omega\). In the first regime, characterized by a rapid reduction in inelastic spectral weight with a time constant of approximately 65 fs, the electron-hole plasma efficiently screens the core exciton. This screening diminishes exciton-phonon coupling and suppresses JT distortions, resulting in dynamics dominated by electronic processes. In the second regime, associated with larger detuning, effective scattering times fall below 2 fs, where local structural distortions become negligible. Here, thermally induced changes in the interplanar distances of the graphite lattice dominate the dynamics.

These results provide a detailed temporal-energy map of the evolution of spectral features, elucidating the transition from electronically driven to thermally activated processes. While the present investigation focuses on graphite, the insights gained are broadly applicable to a wide class of two-dimensional materials where poor screening and strong exciton-phonon coupling play a decisive role. This work thus establishes a framework for understanding ultrafast dynamics in low-dimensional systems and underscores the power of time-resolved RIXS for probing non-equilibrium phenomena in complex quantum materials with high temporal and energy resolution.

\section{Acknowledgments}
We acknowledge the assistance of the staff at FERMI during the beamtimes 20209081 and 20214052. 
\section{Author Contributions}
M. M., R.B., A.C., S.L. carried out the experiment, collected the data, and contributed to the preliminary analysis of the data. The theoretical investigation was conducted by D.N.. M.M., B.V., E.B. performed the data analysis. M.M., A.C. were responsible for the MagneDyn endstation. P.R., E.M.A.,L.G., D.G. and F.S. optimized the accelerator and provided the FEL beam. M.Ma., A.S., M.Z. optimized the optical transport of the MagneDyn beamline and characterized the FEL pulses. A.C., A.D. and P.S. optimized the optical laser system. M.M. wrote the manuscript, which all authors discussed. M.M. proposed and led the project.

\bibliography{exiton_NC}
\end{document}